# Formation of aeolian ripples and sand sorting

**Edgar Manukyan[a] and Leonid Prigozhin[b]**

Department of Solar Energy and Environmental Physics, Blaustein Institutes for Desert Research, Ben Gurion University of the Negev, Sede Boqer Campus, 84990 Israel

We present a continuous model capable of demonstrating some salient features of aeolian sand ripples: the realistic asymmetric ripple shape, coarsening of ripple field at the nonlinear stage of ripple growth, saturation of ripple growth for homogeneous sand, typical size segregation of sand and formation of armoring layers of coarse particles on ripple crests and windward slopes if sand is inhomogeneous.

PACS number(s): 45.70.Qj, 45.70.Mg, 92.60.Gn

## I. INTRODUCTION

Aeolian ripples form regular patterns on sand beaches and desert floors and indicate instability of flat sand surfaces under the wind-induced transport of sand grains. The ripples are oriented perpendicularly to the wind direction and have asymmetric cross sections: their concave lee (downwind) slopes are steeper than slightly convex stoss (upwind) slopes but usually do not reach the sand angle of repose [1,2]. The flattened crests of sand ripples often end with a brink.

[a] Present address: Institute of Geophysics, ETH Zurich, Zurich, Switzerland. E-mail: edgar@aug.ig.erdw.ethz.ch

[b] Corresponding author. E-mail: leonid@math.bgu.ac.il

To analyze the mechanics of sand transport by wind Bagnold [3] proposed to distinguish two types of sand grain movement: saltation and creep. Anderson [1] distinguished also reptation of sand particles. Saltating grains move by long jumps which end in a high-energy impact with the surface. These impacts take place at almost uniform shallow angles of descent varying from $9^o$ to $15^o$ to horizontal [4]. At an impact the saltating grain loses part of its kinetic energy but usually rebounds high enough to be accelerated by the wind again and continues its saltation. Each impact can also cause ejection from the sand bed of several low energy grains which can make short jumps, mostly downwind, and/or roll a little upon the bed surface before they are trapped into the bed again. These are reptating and creeping grains, respectively, but for brevity we will call all low-energy ejecta "reptating grains". Bagnold's explanation of the flat sand surface instability, arising whenever the wind is able to drive sand grains into saltation, was based on the observation that the rate of saltation impacts is higher on the windward than on the lee slopes of any small undulation of bed surface. This explanation remains the basis of modern understanding of the initial stage of flat surface instability, although Bagnold's hypothesis that ripple wavelength is equal to the characteristic length of saltation jump has been challenged by several researchers (see [1, 2]). It seems now commonly accepted that saltation trajectories are widely distributed and, typically, many times longer than the initial ripple wavelength; the ripple wavelength is not constant but grows with time; and the essential physics lies in the variation of reptation flux. The saltating grains gain their momentum from the wind, supply the energy necessary for reptation, and decelerate the wind near the bed surface. As the dynamic equilibrium between the amount of saltating particles and wind shear stress establishes, the probability of direct entrainment of the bed particles into motion by wind becomes small.

Formation of aeolian ripples is strongly influenced by size sorting of sand. Ripples composed of homogeneous sand are small and flat because the surface grains are destabilized by saltation impacts and carried away by the wind, stronger on the ripple crests; these grains may be deposited in troughs [3]. The final wavelength of such ripples, measured after a long period of a unidirectional wind, was correlated with the grain size [5]. Ripples made of inhomogeneous sand become much bigger because coarser grains form an armoring layer on the ripple crests and stoss slopes; significantly less crude particles are found in the troughs. The armoring layer may be very thin, almost a monolayer, see [5], or becomes thick near the ripple crest (see, e.g., [3], Fig. 51). Probably, the inhomogeneous sand ripples continue growing till the wind starts to carry away also coarse grains from their crests. Contrary to ordinary ripples, no significant correlation between the wavelength and the mean grain size has been observed for megaripples [5]. Aeolian ripples larger than those commonly found in fine sands were variously termed ridges, granule ripples, megaripples or gravel ripples; such ripples were studied at various locations on Earth (see, e.g., [2,3,5,6]) and, recently, also on Mars [7,8].

Bagnold has postulated the following conditions necessary for megaripple growth: (i) availability of coarse grains with diameters 3-7 times larger than the mean diameter of grains in saltation; (ii) a constant supply of fine sand in saltation to sustain forward motion of coarse grains by creep; and (iii) wind velocity below the saltation threshold of the coarse grains.

Formation and stabilization of ripples is observed on a time scale of minutes and hours, respectively, but, according to Bagnold [3], it takes decades or centuries to form huge megaripples. Sharp, however, noted that, provided a suitable supply of coarse grains is available, it might take only weeks of sufficiently strong wind to form good granule ripples at the Kelso Dunes in California [2]; the coarse grains in this case composed 50–80 percent of the surface material. Wind tunnel experiments [6], performed with sand collected from granule ripples, showed that, starting with a flat bed, usual ripples form, coarsen and gradually turn into megaripples.

In [1], Anderson modified Bagnold's model for aeolian ripples: the key role in ripple formation was given to reptation jumps; their distribution being described by the "splash function" [9]. Linear stability analysis of this model yields that the initial ripple wavelength is determined by, and is several times larger than, the mean length of reptation. The model, however, demonstrates unrealistic behavior on the nonlinear stage of ripple growth which begins very early; it is, therefore, unable to describe the progressive increase of ripple wavelength observed in the field and wind tunnel experiments. It was supposed [10] that the model can be improved by accounting for possible rolling of reptating particles down the slopes of sand surface.

This idea was used by Landry and Werner in their cellular automata model [11], where saltation impacts not only compelled sand bed grains to jump a short distance downwind: endowing the grains with some initial pseudo-momentum enabled reptating particles to roll after landing upon the bed surface. The model was able to reproduce a non-trivial mechanism of ripple field coarsening via the soliton-like interactions with only partial mass exchange between ripples.

To account for rolling and trapping of reptating particles on the bed surface in a continuous model, Anderson's model can be supplemented by the BCRE (Bouchaud, Cates, Ravi Prakash, Edwards) equations [12] for surface flow of sand and sand surface evolution. Such an approach has been employed in [13] (see also [14]) to study the linear stability of sand bed under saltation. Another continuous model for sand ripples has been proposed in [15], where BCRE equations were simplified (see also [16]) and simulation extended to the nonlinear stage of ripple growth. The latter model was able to simulate realistic shapes of aeolian ripples and ripple coarsening.

Although the key role of sand inhomogeneity is well known, existing mathematical models of aeolian ripples were, with only a few exceptions, written for homogeneous sand. One of these exceptions is the stochastic cellular automaton model for bi-disperse sand [17], where grading resulted from the difference in ejected grain trajectories for crude and fine particles. Grains of both sizes were allowed to saltate which can be justified for the considered mixture: the diameter ratio for crude and fine particles was only 1.39, much smaller than the ratio 3-7 supposed to be necessary for formation of megaripples [3]. To obtain a realistic sand ripple shape and segregation pattern using this model the authors had to introduce significant wind compression over the ripple crests by setting a "ceiling height" for the wind flow at about one wavelength above the mean bed surface.

In a discrete bi-disperse model of aeolian ripples [18], Makse considered different types of segregation and claimed the most important of them is often the percolation of fines through the layer of rolling (reptating) particles towards the bed surface. However, this mechanism of segregation is hardly relevant to formation of sand ripples because the rolling layer is very thin: its effective thickness is less than one particle diameter (see, e.g., the close-up video [19]). Similarly to [17], Makse assumes both types of particles saltate, which means the diameter ratio is small.

A continuous mathematical model for a two-size mixture of grains was recently introduced by Yizhaq ([20], see also [21]). The model takes into account the variability of saltation flux caused, as are the changes in reptation flux, by variations of impact rate due to surface undulations but at much further upwind locations, at a distance of the order of mean length of saltation jump. Linear stability analysis of this model shows the presence of two maxima in the growth rate of unstable modes: one of them is attributed to standard ripples, the other, with a wavelength several times longer, to megaripples. This model, however, does not account for sand segregation and describes megaripples not as the overgrown (because of sorting and armoring effect) usual ripples but as a separate aeolian form between ripples and dunes.

We conclude that existing mathematical models of aeolian sand ripples either ignore size segregation of sand or are not applicable for the size ratios typical of sand in natural ripples and, especially, megaripples. The general physical picture of ripple formation, presented by Bagnold in his classical work on aeolian sand transport [3], should possibly be modified mainly in only one respect concerning the ripple wavelength. Our work is an attempt to derive a continuous mathematical model of aeolian ripples in accordance with Bagnold's and Anderson's views but also employing a simplified BCRE-type model [16] for the surface flow of sand, describing exchange between sand bed and saltation cloud similarly to [22] and, last but not least, accounting for inhomogeneous composition of sand and size segregation.

## II. MATHEMATICAL MODEL

The model is written for a bi-disperse mixture of fine and coarse grains. Impacts of saltating particles destabilize sand grains from the bed surface layer and make them to reptate. The wind shear velocity is assumed below the threshold necessary to drive the coarse particles into saltation, so only small particles are able to saltate. The equation of sand balance on the bed surface can be written as

$$\frac{\partial h}{\partial t} = -\Phi_s^f + \Gamma_r^f - E_r^f + \Gamma_r^c - E_r^c, \qquad (1)$$

where $h(x,t)$ is the bed surface, $t$ is time, $x$ is the coordinate along the wind direction, the lateral coordinate is ignored for simplicity, indices $s$ and $r$ correspond to saltation and reptation, respectively, and indices $f$ and $c$ denote fine and coarse particles, respectively. $E_r^f$ and $E_r^c$ are the rates with which particles from the surface layer are expelled by saltation impacts into reptation, $\Gamma_r^f$ and $\Gamma_r^c$ are the rates with which reptating particles are stopped and incorporated into the bulk, $\Phi_s^f$ is the rate of exchange between the saltating population of fines and sand bed.

The ejection rate of sand particles from the surface layer should be proportional to saltation bombardment intensity. If $q_s$ is the flux of saltating particles and $l_{salt}$ is the mean length of saltation jump, the intensity of bombardment by saltating grains is proportional to $\frac{q_s}{l_{salt}}$. On the other hand, if the saltating grains strike the bed surface at the angle $\gamma$ to horizontal and the bed is inclined at an angle $\alpha$ (see Fig. 1), the incident angle of attack is $\theta = \alpha + \gamma$ and the bombardment intensity should be proportional to $\sin\theta$ (or $\frac{\sin\theta}{\sin\gamma}$ for a given $\gamma$).

This is true for stoss slopes and also the lee slopes that are not too steep. If $\theta \leq 0$, the corresponding part of the surface is in shadow and unreachable by saltating grains. Furthermore, shadowing is a non-local phenomenon and we assume the part of the surface is in shadow also if saltating grains approaching it by the straight trajectories at the angle $\gamma$ to horizontal strike the bed surface somewhere upwind first. In shadow regions the ejection rate for both particle types is set to zero. Introducing the shadow indicator function $S_\gamma(x,t)$ equal to zero if the point $(x, h(x,t))$ is in shadow and to one otherwise, we postulate the erosion rates

$$E_r^f = v^f S_\gamma \frac{q_s}{l_{salt}} \frac{\sin\theta}{\sin\gamma} \varphi(\tilde{\kappa}^f), \qquad E_r^c = v^c S_\gamma \frac{q_s}{l_{salt}} \frac{\sin\theta}{\sin\gamma} \left(1 - \varphi(\tilde{\kappa}^f)\right), \tag{2}$$

where $v^f$ and $v^c$ are dimensionless coefficients and $\varphi(\tilde{\kappa}^f)$ is the "shielding factor". This factor describes shielding of fine particles by the coarse ones and depends on the surface concentration of fine particles, $\tilde{\kappa}^f$, and the diameter ratio (we omit the latter dependence for notation simplicity). When there are no fine particles on the sand bed surface, $E_r^f$ should be zero. Similarly, $E_r^c$ has to be zero wherever there are only fine particles on the surface. Thus, the shielding factor should satisfy the conditions $\varphi(0)=0$, $\varphi(1)=1$. In our simulations the shielding factor was chosen as the size-weighted surface concentration,

$$\varphi\left(\tilde{\kappa}^f\right) = \frac{\tilde{\kappa}^f d^f}{\tilde{\kappa}^f d^f + \left(1-\tilde{\kappa}^f\right)d^c}, \tag{3}$$

where $d^f$ and $d^c$ are diameters of fine and crude particles, respectively.

A reptating particle ejected by an impact at a point $\tilde{x}$ makes a jump and lands on the bed surface at a point $x$ with the probability density given by the splash function, $p = p_h(\tilde{x},x)$, introduced by Ungar and Haff [9]. We assume that reptating particles lose most of their momentum in collision with the rough bed surface but do not stop immediately upon landing and may roll upon the bed surface, usually not far from the landing point, before they get trapped. Rolling of these particles is described by modified quasi-stationary BCRE equations (see [15,16])

$$\frac{\partial}{\partial x}\left(V_r^f R^f\right) = Q^f\left[E_r^f\right] - \Gamma_r^f, \qquad \frac{\partial}{\partial x}\left(V_r^c R^c\right) = Q^c\left[E_r^c\right] - \Gamma_r^c, \tag{4}$$

where $V_r^f$, $V_r^c$ are the horizontal projections of particle velocities and $R^f$, $R^c$ are the equivalent partial thicknesses of rolling layers, so that $V_r^f R^f$ and $V_r^c R^c$ are horizontal projections of the fluxes of rolling particles; $Q^f\left[E_r^f\right]$ and $Q^c\left[E_r^c\right]$ are the rates of precipitation of particles ejected from the bed by saltation impacts (see [1,15]):

$$Q^f\left[E_r^f\right] = \int_{-\infty}^{\infty} E_r^f(\tilde{x},t) p_h^f(\tilde{x},x) d\tilde{x}, \qquad Q^c\left[E_r^c\right] = \int_{-\infty}^{\infty} E_r^c(\tilde{x},t) p_h^c(\tilde{x},x) d\tilde{x}. \tag{5}$$

We assume that rolling of destabilized grains upon the sand surface is due to the gravity force and wind drag; our simple steady-state velocity approximation is based on the assumption of equilibrium between these forces and an effective friction force caused by collisions of a particle with the bed grains and proportional to its velocity. The gravity force tends to carry a rolling particle downslope; the steeper the slope, the stronger is this force. The wind drag acts in the direction of the wind; the higher the shear velocity, the stronger is the force. As the simplest relations for rolling particle velocities we assume

$$V_r^f = -\mu^f \frac{\partial h}{\partial x} + \omega^f u_s^*, \qquad V_r^c = -\mu^c \frac{\partial h}{\partial x} + \omega^c u_s^*, \qquad (6)$$

where $\mu^f$ and $\mu^c$ are constant mobilities of particles, $\omega^f$ and $\omega^c$ characterize the drag forces, $u_s^*$ is the wind shear velocity in the saltation layer near the bed surface. We emphasize that in our model wind drag is unable to remove a particle from the sand bed: the drag influences only the rolling (creep) of reptating particles until they are captured into the bulk again.

Rolling particles never form a thick layer on the surface during the ripple growth and it can be assumed that, for a fixed bed surface incline, the rate of rolling-to-steady-state transition $\Gamma$ is proportional to the amount of rolling particles on the surface, $R$. Furthermore, the steeper the free surface, the lower is the stopping rate. For slopes steeper than the sand angle of repose, $\alpha_r \approx 30°$, rolling sand grains do not stop at all. Following [15,16] we assume, as a simple but physically reasonable approximation for subcritical slopes, that $\Gamma$ is also proportional to $\left(\tan^2 \alpha_r - |\nabla h|^2\right)$ and thus obtain

$$\Gamma_r^f = \gamma^f R^f \left(1 - \frac{|\nabla h|^2}{\tan^2 \alpha_r}\right), \qquad \Gamma_r^c = \gamma^c R^c \left(1 - \frac{|\nabla h|^2}{\tan^2 \alpha_r}\right), \qquad (7)$$

where $\gamma^f$ and $\gamma^c$ characterize trapping and immobilization rates of particles upon a horizontal surface. These rates depend on the composition of surface layer,

$$\gamma^f = \gamma_0 \psi^f\left(\tilde{\kappa}^f\right), \qquad \gamma^c = \gamma_0 \psi^c\left(\tilde{\kappa}^f\right), \tag{8}$$

where $\gamma_0$ has the dimension of frequency and is the trapping rate for particles rolling upon a horizontal bed of particles of the same size; $\psi^f\left(\tilde{\kappa}^f\right)$ and $\psi^c\left(\tilde{\kappa}^f\right)$ are "segregation functions". If a particle rolls upon a bed of larger particles, it is stopped more quickly than on the bed consisting of smaller grains. Thus, if the surface concentration of fines decreases, the stopping rate of both fine and coarse particles should increase. On the other hand, the higher is the diameter ratio of particles the more pronounced is segregation. Finally, we need to satisfy the condition $\psi^f(1) = \psi^c(0) = 1$. Taking these simple arguments into account, we define the segregation functions as follows,

$$\psi^f\left(\tilde{\kappa}^f\right) = 1 + k_\Gamma \left(\frac{d_c}{d_f} - 1\right)\left(1 - \tilde{\kappa}^f\right), \qquad \psi^c\left(\tilde{\kappa}^f\right) = \frac{1 + k_\Gamma \left(\frac{d_c}{d_f} - 1\right)\left(1 - \tilde{\kappa}^f\right)}{1 + k_\Gamma \left(\frac{d_c}{d_f} - 1\right)}, \tag{9}$$

where $k_\Gamma$ is a dimensionless segregation coefficient.

Exchange of fines between the sand bed and saltating cloud occurs in our model wherever saltation flux is not in equilibrium with the local surface shear stress. Even if the flux is the equilibrium one for the incoming wind blowing above flat bed surface, it becomes locally under- or oversaturated due to the surface shear velocity variations caused by surface undulation. The equilibrium sand flux is often calculated by means of the Lettau and Lettau formula [23]; for volumetric flux it reads

$$q^{eq}(u^*) = C_L \frac{\rho_{air}}{\rho_{sand} g} u^{*2} (u^* - u_t^*)_+ \qquad (10)$$

where $C_L = 4.1$, $\rho_{air}$ and $\rho_{sand}$ are the densities of air and sand, respectively, $g$ the acceleration of gravity, $u_t^*$ is the threshold shear velocity necessary to keep saltation motion, $u^*$ is the shear stress above the saltation layer, and $u_+ = \max(u, 0)$. Since we are going to model only a small part of the ripple field, we assume the flux of sand transported by wind through this area is determined by external conditions. Most of this sand is transported by saltation and we will make no difference between the total sand flux and saltation flux, $q_s$. If the wind is locally oversaturated, sand is deposited onto the sand bed and we approximate the exchange rate as

$$\Phi_s^f = \frac{q^{eq}(u^*) - q_s}{l_{sat}} \quad \text{if} \quad q_s > q^{eq}(u^*). \qquad (11)$$

Here $l_{sat}$ is the "saturation length" used (see [22,24,25]) to describe the relaxation of sand flux towards equilibrium and, as in [26], we linearized the original formula [22].

Otherwise, if the wind is locally undersaturated, the fines from the bed are winnowed away with the rate, we assume, proportional to their surface concentration:

$$\Phi_s^f = \tilde{\kappa}^f \frac{q^{eq}(u^*) - q_s}{l_{sat}} \quad \text{if} \quad q_s < q^{eq}(u^*). \qquad (12)$$

Due to exchange between the sand bed and sand transported by wind above and upon the bed surface, the composition of surface layer changes. To simulate this, we assume that bombardment by saltating grains makes the surface layer agitated and mixing in this layer becomes possible. In our model the mixing is described by diffusion,

$$\frac{\partial \kappa^f}{\partial t} = \frac{\partial}{\partial y}\left(D(x,t,h(x,t)-y)\frac{\partial \kappa^f}{\partial y}\right), \qquad -\infty < y < h(x,t), \tag{13}$$

where $\kappa^f(x,y,t)$ is the concentration of fine particles in the sand bed, $y$ is the vertical coordinate, diffusion in the $x$ direction is neglected and, of course, $\kappa^f(x,h(x,t),t)$ is the surface concentration of fines denoted above as $\tilde{\kappa}^f(x,t)$. The diffusion coefficient $D$ is proportional to the bombardment intensity, has the same dimension as saltation flux, and should decay quickly with the distance from the bed surface. We set

$$D = D_0 S_\gamma \frac{q_s}{l_{salt}} \frac{\sin\theta}{\sin\gamma}(\delta_D - [h-y])_+ \tag{14}$$

where $\delta_D$ is the thickness of the diffusion layer and $D_0$ is a dimensionless constant.

To derive the boundary condition at the bed surface, let us consider the cumulative distribution of fine sand along the $x$ axis, $m^f(x,t) = \int_{-\infty}^{h(x,t)} \kappa^f(y,x,t)dy$. Since the diffusion in $x$ direction is neglected,

$$\frac{\partial m^f}{\partial t} = -\Phi_s^f + G_r^f - E_r^f. \tag{15}$$

Making use of the diffusion equation (13) we also obtain

$$\frac{\partial m^f}{\partial t} = \frac{\partial h}{\partial t}\kappa^f\bigg|_{y=h} + \int_{-\infty}^{h}\frac{\partial \kappa^f}{\partial t}dy = \frac{\partial h}{\partial t}\kappa^f\bigg|_{y=h} + D(x,t,0)\frac{\partial \kappa^f}{\partial y}\bigg|_{y=h}$$

and, taking (1) and (15) into account, arrive at the boundary condition

$$D(x,t,0)\frac{\partial \kappa^f}{\partial y} + \left(-\Phi_s^f + \Gamma_r^f - E_r^f + \Gamma_r^c - E_r^c\right)\kappa^f = -\Phi_s^f + \Gamma_r^f - E_r^f, \quad y = h(x,t) \tag{16}$$

To complete the model we need yet to specify the splash functions, $p_h^f$ and $p_h^c$, and describe spatial variations of the shear velocity for a given relief.

Although not much is known about the reptation jump distribution, previous studies (see, e.g., [1]) suggest that the system is not too sensitive to details of this function behavior and an approximation sufficient for qualitative simulation can be obtained by combining existing experimental data and simple physical arguments. Here we employ the one-dimensional approximation of splash function for fine particles by the density of normal distribution derived in [15] using experimental data by Willets and Rice [4]. Namely, we assume that the jumps of fine reptating particles ejected from the bed at the point $x_1$ are distributed approximately as

$$p_h^f(x_1,x_2) = \frac{1}{\sigma_r^f \sqrt{2\pi}} \exp\left(-\frac{1}{2}\left[\frac{x_2 - x_1 - l_r^f (\partial h/\partial x)|_{x_1}}{\sigma_r^f}\right]^2\right). \tag{17}$$

Here the mean jump depends on the surface slope, $l_r^f\left(\frac{\partial h}{\partial x}\right) \approx L\left(1 - 2.01\frac{\partial h}{\partial x}\right)$; the standard deviation $\sigma_r^f \approx 1.25L$; and the mean jump of a fine reptating particle on the horizontal sand surface, $L = l_r^f(0)$, is chosen below as the unit of length. The reptation jump of a crude particle hit by a small saltating grain must be shorter than the jump of a fine one; we assume a similar distribution with $l_r^c = l_r^f k_r$, $\sigma_r^c = \sigma_r^f k_r$ and the proportionality coefficient depending on the diameter ratio: $k_r = k_r(d^c/d^f) < 1$.

Calculating local variations of surface shear stress and velocity in a turbulent flow over a given relief is difficult. The asymptotic solution [27] for turbulent flow over an isolated low hill has been refined in a series of works and employed for simulation of sand dunes in [24]; it yields $\tau = \tau_0(1 + \hat{\tau})$, where $\tau_0$ is a given shear stress and

$$\hat{\tau}(x,t) = A\left(\frac{1}{\pi}\int_{-\infty}^{\infty}\frac{1}{x-\xi}\frac{\partial h}{\partial x}(\xi,t)d\xi + B\frac{\partial h}{\partial x}(x,t)\right). \tag{18}$$

Here $A, B$ are positive, depend weakly on the ratio of characteristic hill length to the roughness length of the surface, and can be taken constants as an approximation. The integral term in (18) is the Hilbert transform of $\partial h/\partial x$ and should be understood in the Cauchy principle value sense. Important features of this solution are complete scale invariance expected in the fully turbulent regime (the same speed-up for small and large perturbations of the same shape) and the characteristic upwind shift of shear stress perturbations in respect to surface variations. The influence of parameters $A, B$ on the isolated dune shape is discussed in [26]; a similar expression for the shear stress in a turbulent flow of liquid over a rippled surface has been recently derived in [28]. We also employ the solution (18) in our model but, since the periodic boundary conditions are assumed, we use an equivalent expression for Hilbert transform of periodic functions (see, e.g. [29]):

$$\frac{1}{\pi}\int_{-\infty}^{\infty}\frac{f(\xi)}{x-\xi}d\xi = \frac{1}{L_x}\int_{0}^{L_x}f(\xi)\cot\left(\frac{\pi(x-\xi)}{L_x}\right)d\xi. \tag{19}$$

Here $L_x$ is the period (the length of a part of the sand bed considered in our model). This gives

$$\hat{\tau}(x,t) = A\left(\frac{1}{L_x}\int_{0}^{L_x}\frac{\partial h}{\partial x}(\xi,t)\cot\left(\frac{\pi(x-\xi)}{L_x}\right)d\xi + B\frac{\partial h}{\partial x}(x,t)\right). \tag{20}$$

We note that, although this solution has qualitatively correct scaling and upwind shift, it should yet be verified for periodic reliefs and does not take into account the feedback effect of saltation upon the air flow. The feedback can, probably, influence at least the values of parameters $A, B$.

The shear velocity above the saltation layer is determined as $u^* = \sqrt{\tau/\rho_{air}}$. We define also the nominal value of shear velocity, $u_0^* = \sqrt{\tau_0/\rho_{air}}$, and the corresponding equilibrium volumetric sand flux, $q_0^{eq} = q^{eq}(u_0^*)$, determined by (10). The shear velocity $u_s^*$ (close to the bed surface inside the saltation layer) is decreased by saltation drag and, in equilibrium above the flat sand surface, becomes close to the threshold velocity $u_t^*$. To account for its variation due to uneven relief in Eq. (6), we assume $u_s^*$ is proportional to $u^*$ but scaled as $u_t^*$ and not as $u_0^*$, i.e., $u_s^* = \dfrac{u_t^*}{u_0^*} u^*$.

## III. SCALING

We now rescale the model variables and rewrite the equations in dimensionless form. Let

$$\{x', y', h'\} = \frac{1}{L}\{x, y, h\}, \qquad t' = \frac{t}{T}, \qquad R^{f\prime} = \frac{R^f}{R_0^f}, \qquad R^{c\prime} = \frac{R^c}{R_0^c},$$

$$\{E_r^{f\prime}, E_r^{c\prime}, \Gamma_r^{f\prime}, \Gamma_r^{c\prime}, Q^{f\prime}, Q^{c\prime}, \Phi_s^{f\prime}\} = \frac{T}{L}\{E_r^f, E_r^c, \Gamma_r^f, \Gamma_r^c, Q^f, Q^c, \Phi_s^f\},$$

$$q_s' = \frac{q_s}{q_0^{eq}}, \qquad u^{*\prime} = \frac{u^*}{u_0^*}, \qquad \tau' = \frac{\tau}{\tau_0}, \qquad \kappa' = \kappa.$$

Here the unit of length is the mean reptation jump of fine particles upon a flat horizontal surface, $L = l_r^f(0)$; the unit of time is chosen as the time needed to the equilibrium flux of saltation particles $q_0^{eq}$ to expel into reptation a layer of thickness $L$ from the flat horizontal bed of fine particles, $T = \dfrac{L l_{salt}}{v^f q_0^{eq}}$. Characteristic thicknesses of rolling layers are chosen as the equilibrium ones ensuring the balance between sand ejection and immobilization on flat horizontal bed surface if the saltation flux is $q_0^{eq}$ and the bed consists of only one type of particles:

$$R_0^f = \frac{v^f q_0^{eq}}{l_{salt} \gamma_0}, \qquad R_0^c = \frac{v^c q_0^{eq}}{l_{salt} \gamma_0}.$$

The scaling of balance equations yields (the primes are omitted):

$$\frac{\partial h}{\partial t} = -\Phi_s^f + \Gamma_r^f - E_r^f + \Gamma_r^c - E_r^c \tag{21}$$

$$\frac{\partial}{\partial x}\left(\left\{-\frac{\mu^f}{L\gamma_0}\frac{\partial h}{\partial x} + \frac{\omega^f u_t^*}{L\gamma_0}u^*\right\}R^f\right) = Q^f\left[E_r^f\right] - \Gamma_r^f \tag{22}$$

$$\frac{\partial}{\partial x}\left(\left\{-\frac{\mu^c}{L\gamma_0}\frac{v^c}{v^f}\frac{\partial h}{\partial x} + \frac{v^c}{v^f}\frac{\omega^c u_t^*}{L\gamma_0}u^*\right\}R^c\right) = Q^c\left[E_r^c\right] - \Gamma_r^c \tag{23}$$

where

$$E_r^f = S_\gamma q_s \frac{\sin\theta}{\sin\gamma}\varphi(\tilde{\kappa}^f), \qquad E_r^c = S_\gamma \frac{v^c}{v^f} q_s \frac{\sin\theta}{\sin\gamma}\left(1-\varphi(\tilde{\kappa}^f)\right), \tag{24}$$

$$Q^f\left[E_r^f\right] = \int_{-\infty}^{\infty} E_r^f(\tilde{x},t) p_h^f(\tilde{x},x) d\tilde{x}, \qquad Q^c\left[E_r^c\right] = \int_{-\infty}^{\infty} E_r^c(\tilde{x},t) p_h^c(\tilde{x},x) d\tilde{x}, \tag{25}$$

$$\Gamma_r^f = R^f \psi^f(\tilde{\kappa}^f)\left(1-\frac{|\nabla h|^2}{\tan^2\alpha_r}\right), \qquad \Gamma_r^c = \frac{v^c}{v^f} R^c \psi^c(\tilde{\kappa}^f)\left(1-\frac{|\nabla h|^2}{\tan^2\alpha_r}\right), \tag{26}$$

$$\Phi_s^f = \begin{cases} \dfrac{l_{salt}}{v^f l_{sat}}\left(q^{eq}(u^*)-q_s\right) & \text{if } q_s > q^{eq}(u^*), \\ \tilde{\kappa}^f \dfrac{l_{salt}}{v^f l_{sat}}\left(q^{eq}(u^*)-q_s\right) & \text{if } q_s \leq q^{eq}(u^*). \end{cases} \tag{27}$$

The rescaled surface shear stress is $\tau = 1 + \hat{\tau}$, where $\hat{\tau}$ is determined by (20) with the length $L_x$ measured in units of $L$. In the new variables $u^* = \sqrt{\tau}$ and $q^{eq}(u^*) = \frac{(u^* - u_t^*)_+}{1 - u_t^*} u^{*2}$. The diffusion equation (13) and its boundary condition (16)

remain unchanged but the diffusion coefficient is rescaled:

$$D = \frac{D_0}{v^f} S_\gamma q_s \frac{\sin\theta}{\sin\gamma} \left(\frac{\delta_D}{L} - [h-y]\right)_+ \qquad (28)$$

## IV. PARAMETERS

Available experimental data on aeolian transport are scarce and we tried to make a reasonable choice of model parameters guided also by the results of numerical simulations and our limited theoretical understanding of complicated interactions between wind, sand bed, saltating and reptating grains, etc.

Intuitively, it seems the reptation jumps of crude particles should be about $(d^c/d^f)^3$ times shorter than those of fines. On the other hand, this dependence can, possibly, be too strong. Indeed, collision of a saltating grain with the sand bed is not a simple event (see, e.g., [4,30]): usually, the saltating grain enters the bed, emerges out of it some distance away from the point of impact, and compels several sand grains to reptate. We suppose that if, however, a large particle was hit, the saltating grain often rebounds immediately and transfers its momentum mainly to that particle; in our simulations below we assumed $l_r^c = l_r^f k_r, \sigma_r^c = \sigma_r^f k_r$ with $k_r = (d^f/d^c)^2$. We note, however, that using $k_r = (d^f/d^c)^3$ does not change the results significantly. In both cases rolling upon the bed surface has more influence on the reptation of coarse particles expelled from the bed by saltation impacts than their short initial jumps.

We assume for simplicity the mobilities of fine and coarse particles are equal, $\mu^c = \mu^f$, and account for different rolling distances of these particles by employing different trapping rates (26) in Eqs. (22)-(23). This is achieved, first, by introducing a significant segregation coefficient $k_\Gamma = 1$ (see Eq. (9)) and, second, by assuming the wind drag has possibly stronger influence on the crude rolling particles than on the small ones because larger particles, expelled by saltation impacts and rolling in the shear flow upon the bed surface, are protruded further into the wind. The close-up video [19], showing saltation and reptation in a wind tunnel, does, possibly, confirm this assumption. We chose the drag coefficients $\omega^f, \omega^c$ proportional to the size of dragged particles: $\dfrac{\omega^c}{\omega^f} = \dfrac{d^c}{d^f}$.

Exchange of sand between the bed and saltation cloud, accounted for in our model, leads to erosion on ripple crests and deposition in troughs; this flattens the ripples and increases the ripple index, the wavelength to height ratio. We set $\dfrac{\mu^f}{L\gamma_0} = 1.25$, which is less than the value 2 employed in [15], where such exchange was ignored, to obtain a natural index value. We assume that saltation impacts eject the fine particles from the bed surface easier than the crude ones; on the other hand, the fines are less exposed to these impacts. We set the ejection rate coefficients inversely proportional to the size of ejected sand particles, $\dfrac{v^f}{v^c} = \dfrac{d^c}{d^f}$. Following [17,31] we take, as a characteristic value, $u_0^* = 40 \ cm \cdot s^{-1}$ and set $u_t^* = 22 \ cm \cdot s^{-1}$ (which corresponds to $d^f = 0.02$ cm) in dimensional units; the dimensionless values are $u_0^* = 1$ and $u_t^* = 0.55$. Since $\dfrac{\omega^c}{\omega^f} = \dfrac{d^c}{d^f}$ and $\dfrac{v^f}{v^c} = \dfrac{d^c}{d^f}$ we have $\dfrac{\omega^f u_0^*}{l_r^f \gamma_0} = \dfrac{v^c}{v^f} \dfrac{\omega^c u_0^*}{l_r^f \gamma_0}$, see Eqs. (22)-(23); in our simulations the common value of these dimensionless complexes was 0.14 or 0.16. The angle to horizontal at which saltating grains strike the bed surface, $\gamma$, was $11°$ (similar results were obtained for $\gamma$ uniformly distributed between $9°$ and $15°$).

We assumed the characteristic value of $l_{salt}$ is about 50cm, see [32]; $l_{sat}$ is of the order of several meters, see [25,31]; in our simulations we set $l_{salt}/l_{sat} = 0.1$. The mean reptation jump of fine particles, $L = l_r^f(0) = 0.3\,cm$, is about fifteen fine particle diameters. Calculating $q_0^{eq} = 0.056\,cm^2 \cdot s^{-1}$ from Eq. (10), we obtain $T = \dfrac{L l_{salt}}{v^f q_0^{eq}} \approx \dfrac{270}{v^f}$ s. In our simulations we used $v^f = 4$ and small ripples formed during a realistic time, from several minutes to an hour. It may be noted that different authors give different estimates for $l_{sat}$ and $l_{salt}$. Thus, according to Anderson, $l_{salt} \approx 30$ cm, see [1], and Bagnold has found experimentally that $l_{sat} \approx 2.5$m, see [3]. Choosing $l_{salt} \approx 25$cm and keeping the ratio $l_{salt}/l_{sat} = 0.1$ yields Bagnold's estimate for the saturation length. In this case, if we leave unchanged also other dimensionless complexes in our simulations, the only difference would be a twice smaller time scale $T$, which is acceptable for a qualitative model like ours.

The thickness of diffusion layer should be of the order of several fine grain diameters; we took $\delta_D = 0.07\,cm$ and set $D_0 = 0.5$.

For the characteristic ratio of ripple length to sand surface roughness, $10^3 - 10^4$, the parameters in Eq. (20), estimated theoretically in [24,33] without accounting for the saltation feedback on the wind, were $A \approx 5$, $B \approx 1$. We found, however, that without increasing the value of $B$ it is difficult to obtain a realistic armoring layer on the stoss slopes of ripples; in our simulations we used $A = 5$ and $B = 2$.

Since we consider only a small part of the ripple field, the incoming sand flux $q_s$ is determined by external conditions and we neglect its variation in the considered area caused by erosion and deposition of sand. One possible approach is to keep the sand flux constant during the simulation assuming, for example, that we model formation of ripples in a small part of the ripple field where, initially, the surface was made flat artificially.

## V. NUMERICAL APPROXIMATION

The model equations with periodic boundary conditions in $x$ were discretized and solved numerically. We regularized the equation for surface evolution (21) by adding a small diffusion term, $\varepsilon_h \partial^2 h / \partial x^2$ with $\varepsilon_h = 10^{-2}$, and used explicit finite difference scheme. Note that diffusion smoothes the free surface and makes the Hilbert transform of $\partial h / \partial x$ in (20) well defined. The quasistationary BCRE equations for surface flow of sand, (22) and (23), were also regularized, as in [16], by adding diffusion terms, $\varepsilon_R \partial^2 R^f / \partial x^2$ and $\varepsilon_R \partial^2 R^c / \partial x^2$, respectively; $\varepsilon_R = 10^{-2}$. Piecewise linear finite elements were employed to discretize these equations in space. Sufficient resolution in space was obtained with about one thousand equidistant nodes in horizontal direction.

To solve the moving boundary problem (13), (16) we fixed the boundary using the coordinate transform $z = h(x,t) - y$ and arrived at the advection-diffusion problem

$$\frac{\partial \kappa^f}{\partial t} + \left(-\Phi_s^f + \Gamma_r^f - E_r^f + \Gamma_r^c - E_r^c\right)\frac{\partial \kappa^f}{\partial z} = \frac{\partial}{\partial z}\left(D(x,t,z)\frac{\partial \kappa^f}{\partial z}\right), \qquad 0 < z < \infty \quad (29)$$

$$-D(x,t,0)\frac{\partial \kappa^f}{\partial z} + \left(-\Phi_s^f + \Gamma_r^f - E_r^f + \Gamma_r^c - E_r^c\right)\kappa^f = -\Phi_s^f + \Gamma_r^f - E_r^f, \qquad z = 0 \quad (30)$$

which was explicitly discretized in time and solved (at each time step and for each node of a uniform $x$-grid) on a finite interval $0 \leq z \leq L_z$, where $L_z$ was sufficiently large. Since the diffusion coefficient is zero for $z \geq \delta_D$, no boundary condition is needed at $z = L_z$. Here it was necessary to use a total variation diminishing scheme (we used the "superbee" flux limiting method, see [34,35]) to eliminate the approximation viscosity that causes unphysical smearing of concentration variations inside the sand bed.

Initially, the bed had a flat surface slightly perturbed by a random noise; the bed composition was uniform; the bed depth $L_z$ was not more than twice larger than the diffusion length $\delta_D$; and $z$-grid contained only 12 or 16 equidistant nodes. As the ripples formed, non-uniformity of bed composition penetrated deeper into the bed. The sand bed was, therefore, automatically extended when necessary by adding several lower layers with the initial composition of sand; the adaptive mesh generation accelerated our computations. In our longest simulations of inhomogeneous sand ripples ($\approx 7 \cdot 10^5$ time steps) the number of nodes in $z$ direction grew up to two hundreds.

Since the probability density (17) is negligible outside small areas several standard deviations long, computing the integrals in (25) is fast. Finally, the Hilbert transform in (20) is a cyclic convolution efficiently computed by means of the fast Fourier transform.

## VI. SIMULATION RESULTS

*(a) Homogeneous sand.* For sand consisting of the fine particles only, our model is similar to model [15] with two extensions. First, in the new model we account for mass exchange between the bed and saltation cloud; second, drag force is acting on the reptating particles. The initial stage of ripple growth (Fig. 2) is very similar to that in [15]: ripples appear and become asymmetric shortly after the appearance of shadow regions; then coarsening of the ripple pattern takes place. The coarsening does not occur via simple merging of ripples when smaller ripples, moving faster, overtake the larger ones. As it was noted before, see [11,15,36], when an overtaking ripple comes close to a larger one, the trough between them becomes shallow and there appears a "two-headed" ripple. Then, typically, a small ripple, smaller than the overtaking one before the interaction, runs further downwind. Our model reproduces this soliton-like interaction with partial mass exchange; such two-dimensional picture corresponds well to the dynamics of ripple terminations and bifurcations, leading to ripple field coarsening [37], if the dynamics is observed in a cross-section of the ripple field.

As the ripples grow, their velocity and evolution slow down. Previous mathematical models (see, e.g., [11,15,20,38]) could not simulate the ripple wavelength saturation and showed only an increasingly slower, possibly, logarithmic growth in time. This was not completely satisfactory because both field observations and wind tunnel experiments [3,39,31] suggest that ripple growth saturates. In our simulations the saturation of ripple growth was much better reproduced, see Fig. 3, and resulted, as was suggested by Bagnold, from removal of sand by wind from the ripple crests and its deposition in the troughs. The simulated mature ripples have wavelength 9.6 cm and the height about 0.7 cm; their ripple index is thus about 14.

Recently, Andreotti *et al.* [31] presented interesting results of a wind tunnel study of aeolian ripples. First, starting an experiment with a flat sand surface, they observed slowing down of ripple growth and coarsening but were unable to run the experiment long enough to make saturation evident. Therefore they repeated the experiment starting with ripple pattern of the obtained wavelength imprinted upon the sand surface artificially; this allowed them to observe saturation. Their further experiments showed that, under the same conditions, there is a range of stable ripple wavelengths: if the imprinted initial ripple pattern wavelength was changed slightly, the wavelength persisted, although the ripple amplitude could adjust itself before the saturation took place. Significant change of the initial surface wavelength leads to pattern reconstruction via splitting or merging of ripples. These results are nicely reproduced in our simulations, see Figs. 4 and 5, which started with an initially rippled sand surface with the wavelengths two and three times longer, respectively, than the wavelength of saturated ripples formed under the same conditions on the initially flat sand surface.

The influence of drag was not very significant (see Fig. 6); we believe that drag plays more important role (see below) for coarse particles having much shorter reptation jumps.

*(b) Inhomogeneous sand.* Solution of the advection-diffusion equation (29) is time consuming and to model several hours of inhomogeneous sand ripple evolution we needed several days of CPU time. It was, therefore, not possible to model formation of megaripples which needs at least weeks or even months of strong winds. Nevertheless, we were able to follow the initial stage of this process.

Numerical simulations were performed for sand mixtures with diameter ratios $\frac{d^c}{d^f} = 3$ and $\frac{d^c}{d^f} = 5$, consisting in both cases of 90% (volume) of fine particles and 10% of coarse particles, i.e., $\kappa^f(x, y, 0) = 0.9$. We found that for such mixtures switching off the wind drag, acting upon the reptating particles, prevents formation of ripples in our model: the coarse grains remain almost unmovable under saltation strikes and, even if their concentration is small, efficiently stabilize the sand surface. If, however, their reptation is assisted by the wind drag, the ripples form (Figs. 7 and 8). In such a case, after the initial stage of ripple growth, similar to that for homogeneous sand, size segregation gradually develops. The ripple stoss slopes and crests become depleted of fine particles while the concentration of fines in the troughs increases (Fig. 9). On the stoss slopes the armoring layer of crude particles is very thin; however, the whole crests of grown-up ripples contain significant concentration of coarse grains.

Although the ripples at $t \approx 13h$ (Figs. 7, 8) are not large, their growth has slowed down and is controlled by creep of coarse grains on the free surface and diffusion of fines through the surface layer of coarse particles. In comparison to homogeneous sand ripples growing under similar conditions, these ripples are of similar size but less uniform. Since smaller ripples move faster than the large ones, this is a sign that the evolution and coarsening of ripple field are far from finished.

In these examples (Figs. 7, 8) the sand flux was kept constant, $q_s = 1.1 q_0^{eq}$, where $q_0^{eq} = q_0^{eq}(u_0^*)$ is the equilibrium flux determined by formula (10). Other approximate formulas relating the transport rate of blown sand to wind friction velocity and mean grain size have been also used with some success (see, e.g., [40] and the references therein). On the other hand, in our model exchange between the sand bed and saltating grain population is determined locally; it depends not only on the sand flux but also on local surface layer composition and shear velocity, see Eqs. (11)-(12). Given the nominal shear velocity $u_0^*$, for each rippled bed of inhomogeneous sand one can define the equilibrium sand flux $q_{av}^{eq}$, corresponding to zero total mass exchange, i.e., the root of nonlinear equation $\int \Phi[q_s] dx = 0$. We found that such equilibrium-on-average sand flux $q_{av}^{eq}$ depends strongly on the bed surface undulations and sand composition. This seems to be in contradiction with the robust estimates obtained, e.g., using Eq. (10) without taking such details into account. However, our simulations showed that, when the very first small ripples appear on the initially flat surface, the flux $q_{av}^{eq}$ quickly increases and becomes greater than the fixed sand flux $q_s$. The wind becomes undersaturated; fine particles are winnowed away and surface concentration of crudes increases on the stoss slopes and ripple crests.

This acts as a feedback which gradually decreases the difference between $q_{av}^{eq}$ and $q_s$ (see Fig. 10) and also slows down the exchange between bed and saltation cloud. We suppose that if the acting sand flux $q_s$ is not kept constant artificially, as in our simulations of a small part of the sand bed, but determined non-locally by conditions in a much larger area, joint evolution of $q_{av}^{eq}$ and $q_s$ may lead to the equilibrium flux close to $q_0^{eq} = q_0^{eq}(u_0^*)$.

## VII. CONCLUSIONS

Formation of aeolian ripples is influenced by a variety of different factors and, correspondingly, significant variation of ripple forms is found in nature. A realistic model of sand ripples should account for complex interaction of several physical processes: saltation and reptation, turbulent grain-loaded air flow above a given relief, entrainment of sand particles into such flow, and, finally, size segregation of sand accompanying and influencing these processes. Each of the processes continues to be an issue of intensive study and, although general physical understanding of aeolian sand transport has much improved recently, quantitative description remains hardly possible.

Using the simplest mass balance equations and crude empirical constitutive relations, we tried to derive a mathematical model of sand ripples capable of demonstrating the salient features of ripple evolution: the realistic asymmetric ripple shape, coarsening of ripple field, saturation of aeolian ripple growth for homogeneous sand, typical size segregation if sand is inhomogeneous. The proposed bi-disperse model can be considered a simplified scheme of material fluxes; details of this scheme can be changed easily and this makes it a convenient tool for studying the effect of different physical assumptions and mathematical approximations.

Our model accounts for the exchange of sand between the bed and saltation cloud as the mechanism for ripple stabilization suggested by Bagnold [3]: as the ripples grow, more sand is winnowed away from their crests and deposited in the troughs. Numerical simulations confirmed that this model describes saturation of ripple growth for homogeneous sand much better than previous models. In accordance with the wind tunnel experiments [31] we could also reproduce the existence of a range of stable ripple wavelengths.

For inhomogeneous sand having a realistic ratio of coarse to fine particle sizes, sand ripples in our model form only if the reptation of coarse grains is assisted by the wind drag. Size segregation of sand in ripples occurs due to the different modes of fine (saltation and reptation) and crude (reptation) grain motion. Our model accounts for this difference and takes into account also additional segregation mechanism due to the different immobilization rates of reptating fine and coarse grains on the bed surface. Using this model we were able to simulate qualitatively correct segregation pattern and formation of armoring surface layers of crude particles in natural ripples. Although we were not able to perform simulations on the time scale of megaripple formation, modeling the initial stage of ripple growth indicated that for inhomogeneous sand growth of ripples continues further without saturation.

## ACKNOWLEDGMENTS

This work was supported by the Israel Science Foundation (grant N531/06). The paper is a part of EM's M.Sc. project at the Albert Katz International School for Desert Studies (Ben Gurion University of the Negev). LP appreciates discussions with J.W. Barrett, H. Tsoar and H. Yizhaq.

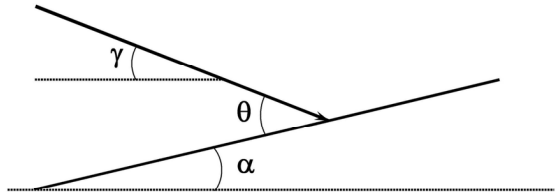

Fig. 1. One-dimensional scheme: saltating grain reaches the bed surface at the angle $\gamma$ to horizontal; the surface is inclined at an angle $\alpha$; $\theta = \alpha + \gamma$ is the incident angle of attack.

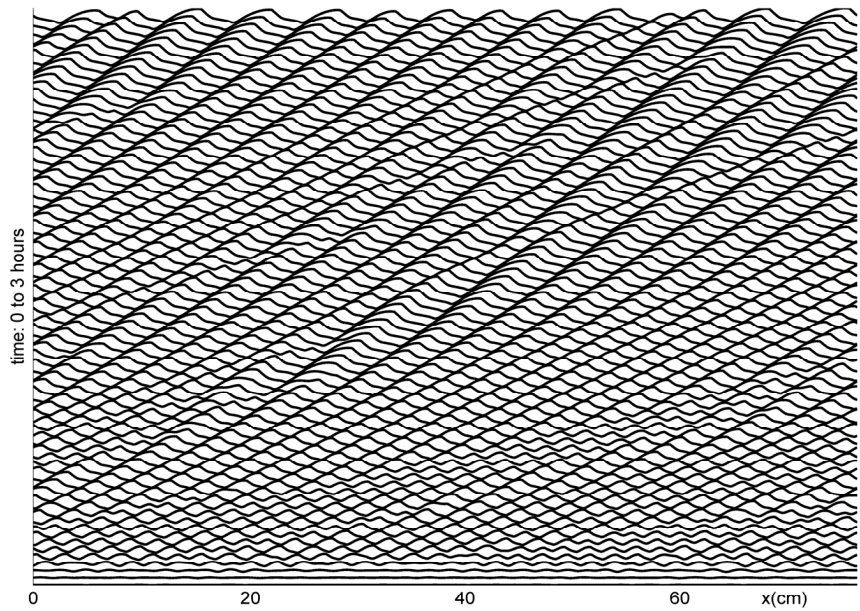

Fig. 2. Homogeneous sand: initial growth and coarsening of sand ripples.

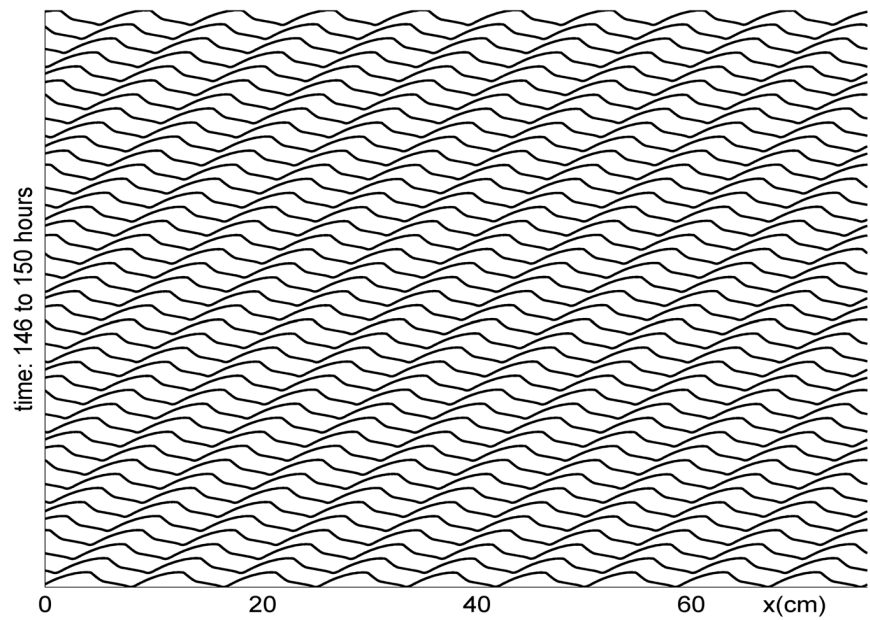

Fig. 3. Homogeneous sand: downwind translation of mature ripples.

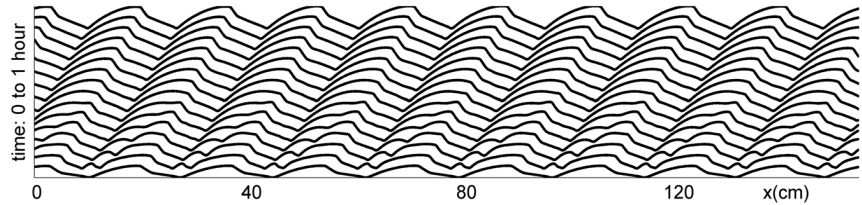

Fig. 4. The mature ripples (see Fig. 3) are stretched 2:1 in x-direction and used as initial sand surface. The ripple height quickly adapts itself to the new ripple wavelength which remains twice longer than in Fig. 3.

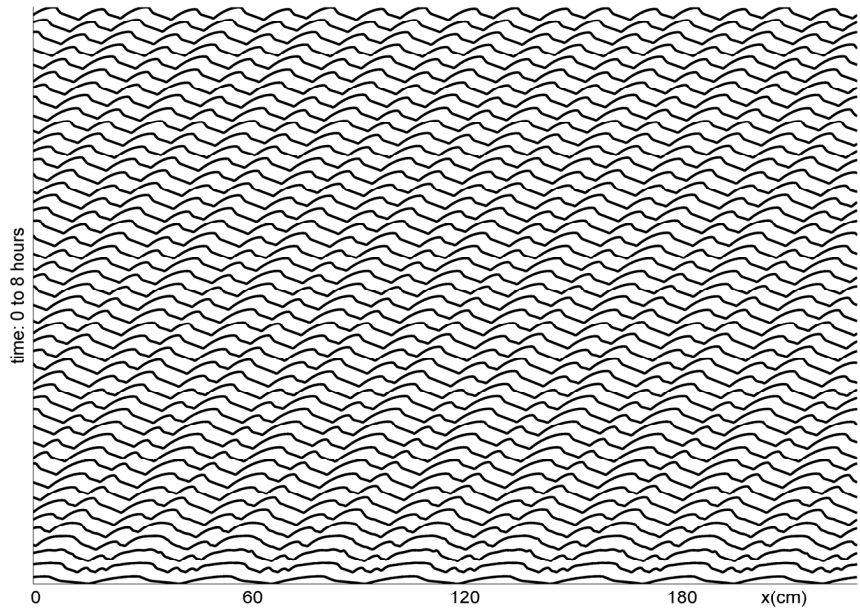

Fig. 5. As in Fig. 4 but the ripples are stretched 3:1 in x-direction. Both the ripple height and wavelength change; the new saturated ripple wavelength is twice shorter than the initial one and 1.5 longer than the wavelength of ripples formed on the initially flat surface.

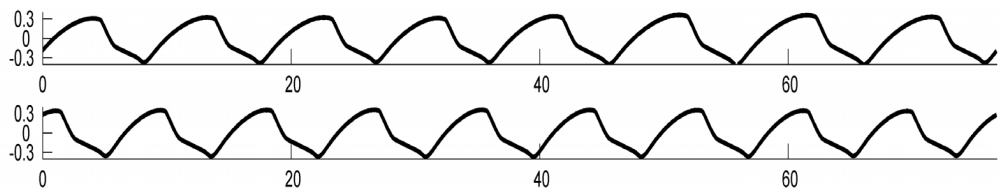

Fig. 6. Simulated mature ripples (homogeneous sand). Upper plot: no drag, $\omega^f = 0$; bottom plot: with drag, $\dfrac{\omega^f u_t^*}{l_r^f \gamma_0} = 0.14$. The lengths are in centimeters.

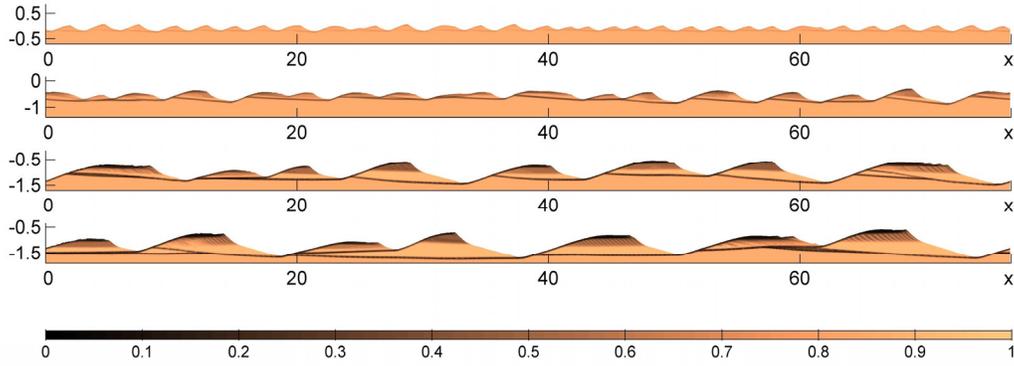

Fig. 7. Ripple growth for bi-dispersed sand mixture containing 10% of crude grains; size ratio $d^c/d^f = 3$; the rescaled drag coefficient $\dfrac{\omega^f u_t^*}{l_r^f \gamma_0} = 0.16$. Times, from top to bottom: 0.3, 1.7, 5.6, 13.5 hours. The sand flux is constant and equal to the oversaturated flux with saturation ratio 1.1 for the flat bed of fine particles. The lengths are in centimeters. The bottom plot shows the color scale for the concentration of fine particles.

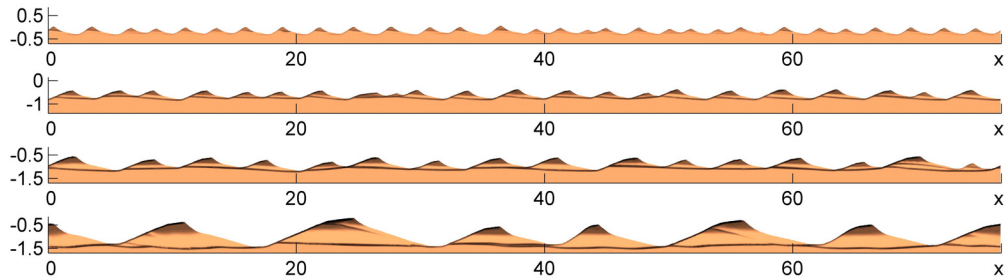

Fig. 8. As in Fig. 7 but for the size ratio $d^c/d^f = 5$ and rescaled drag coefficient $\dfrac{\omega^f u_t^*}{l_r^f \gamma_0} = 0.14$. Times, from top to bottom: 0.3, 1.7, 5.6, 13 hours.

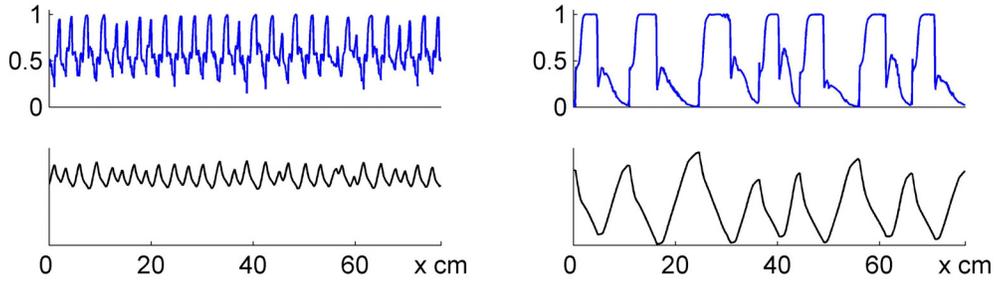

Fig. 9. The surface concentration of fines (top) and rippled bed surface (bottom); $d^c/d^f = 5$. Left: $t = 0.5\,\text{h}$, right: $t = 13\,\text{h}$.

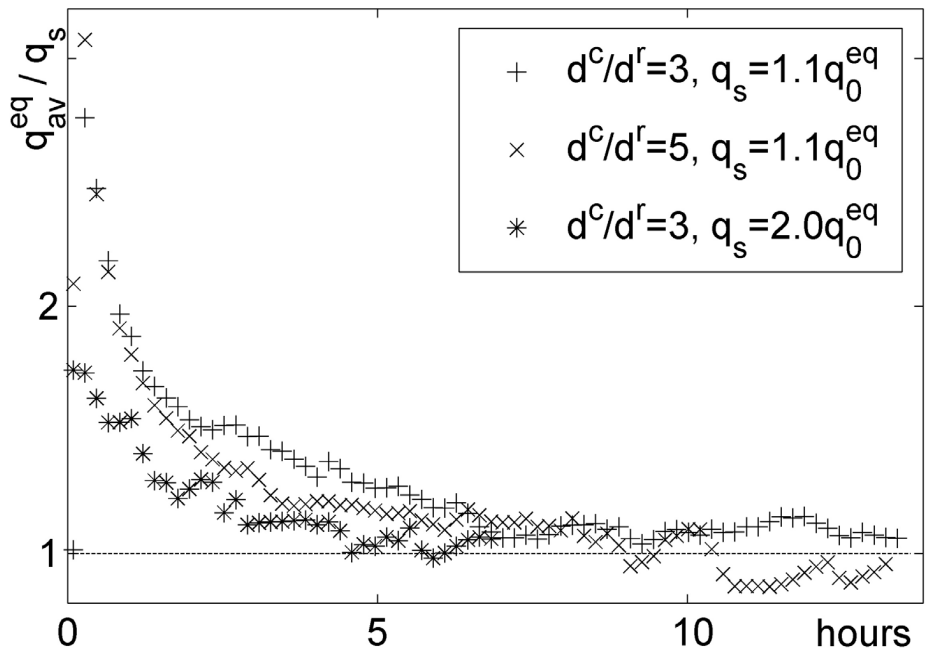

Fig. 10. The equilibrium-on-average sand flux: evolution induced by formation of ripples accompanied by size segregation of sand. As the ripples grow, $q_{av}^{eq}$ becomes close to the acting flux $q_s$.

- 36 -